\let\accentvec\vec
\documentclass[runningheads]{llncs}

\let\vec\accentvec

\usepackage{graphicx}
\usepackage{multirow}
\usepackage{booktabs}
\usepackage{tensor}
\usepackage{amsmath}
\usepackage{amssymb}
\usepackage{stmaryrd}
\usepackage{pbox}
\usepackage{anyfontsize}
\usepackage{url}
\usepackage{algorithm}
\usepackage{algorithmic}
\usepackage{hhline}
\usepackage{makecell}
\usepackage{enumitem}

\begin{document}
\title{Expert Recommendation via Tensor Factorization with Regularizing Hierarchical Topical Relationships}
\titlerunning{Expert Recommendation via Tensor Factorization}

\author{Chaoran Huang\inst{1} \and
Lina Yao\inst{1} \and
Xianzhi Wang\inst{2} \and
Boualem Benatallah\inst{1} \and
Shuai Zhang\inst{1} \and
Manqing Dong\inst{1} }
\authorrunning{C. Huang et al.}
% First names are abbreviated in the running head.
% If there are more than two authors, 'et al.' is used.
%
\institute{UNSW Sydney, NSW 2052, Australia \\ \email{\{chaoran.huang,lina.yao\}@usnw.edu.au}\and
University of Technology Sydney, Broadway, NSW 2007, Australia\\ \email{sandyawang@gmail.com}}

\maketitle

\begin{abstract}
Knowledge acquisition and exchange are generally crucial yet costly for both businesses and individuals, especially when the knowledge concerns various areas. Question Answering Communities offer an opportunity for sharing knowledge at a low cost, where communities users, many of whom are domain experts, can potentially provide high-quality solutions to a given problem. In this paper, we propose a framework for finding experts across multiple collaborative networks. We employ the recent techniques of tree-guided learning (via tensor decomposition), and matrix factorization to explore user expertise from past voted posts.
Tensor decomposition enables to leverage the latent expertise of users, and the posts and related tags help identify the related areas. The final result is an expertise score for every user on every knowledge area. We experiment on Stack Exchange Networks, a set of question answering websites on different topics with a huge group of users and posts. Experiments show our proposed approach produces steady and premium outputs.
\end{abstract}

\section*{Note}
{\footnotesize This article is accepted as full research paper at the 16th International Conference on Service Oriented Computing (ICSOC2018). Hanzhou, China, Nov 12 - Nov. 15, 2018.}

\section{Introduction}
\label{sec:intro}
Question and Answering (Q\&A) websites are gaining momentum as an effective platform for knowledge sharing.
These websites usually have numerous users who continuously contribute.
Many researchers have shown interests in the recommendation issues on these websites such as identifying experts.
Despite the tremendous research efforts on user recommendation, no state-of-the-art algorithms consistently stand out compared with the others.
As the recent work increasingly focuses on domain-specific expertise recommendation, there emerges the research on multi-domain (or cross-domain) recommendation in the ``Stack Exchange (SE) Networks''\footnote{\url{stackexchange.com}} repository.
SE is a network of 98 Q\&A subsites
, all following the same structure. This consistency enables us to expand our approach from one subsites to the all the other subsites on SE.
These subsites cover various disciplines from computer science to even the Ukrainian language.
Take ``Stack Overflow''\footnote{\url{stackoverflow.com}} (SO) as an example( Figure~\ref{fig:seeg}). It is a software-domain-oriented website where users can post and answer questions, or vote up/down to other users' questions and answers. The author of a question (a.k.a., the requester) can mark an answer as accepted and offer a bounty to the answerer.

So far, there are two popular ways to locate experts: collaborative filtering(CF) and content-based recommendation. The former extracts similar people without understanding the contents while the latter focuses on building user profiles based on users' activity history. 
CF relies merely on ratings (e.g., scores in SE networks) and therefore may not well handle sparse Q\&A subsites data, where many questions involve very limited users.
Usually, users can vote on questions, and the vote counts can serve as ratings to the questions.
An earlier work \cite{amatriain2009wisdom} also suggests that the lack of information can be a challenge for recommendation techniques. The work aims to address the data sparsity issue by selectively using the ratings of some experts. This experts presumed by this approach is exactly the same experts we aim to find.
As for content-based approaches, a typical approach (e.g., \cite{Liu2013IEP}) builds user profiles based on user's knowledge scores and user authority in link analysis. The knowledge scores are called reputation in \cite{Liu2013IEP}, which is derived from users' historical question-answering records.
Srba et al. \cite{srba2016stack} point out that some users may maliciously post low-quality content, and those highly active spammers might be taken as experts in a system.
Huna et al. \cite{huna2016exploiting} solve this problem by calculating question and answer difficulties based on three aspects of hints: the numbers of user-owned questions and answers, time difference of the question being posted and answered, average answering time, and score of the answer with the maximum of score among all the answers provided by the answerer. Although these approach may compute user reputation, they also take considerable cost on building user profiles. Matrix Factorization is one method that works on sparse data
% \cite{Koren:2009:MFT:1608565.1608614}
, while matrices can only store two dimensions of data, which is not handy in many applications, where users' attributes can be vital to the identification of experts. Recently tensor-based approaches became popular as an alternative to matrix factorization, made it feasible to handle multi-faceted data\cite{yao2018collaborative}. For example, Ge et al. in \cite{ge2016taper} decompose a (Users, Topics, Experts) tensor for the personalized expert recommendation; Bhargave et al. \cite{bhargava2015and} propose a (User, Location, Activity, Time) tensor decomposition along with correlated matrix to make recommendations based on user preferences. 

\begin{figure}[!h]
\centering
\includegraphics[width=0.6\linewidth]{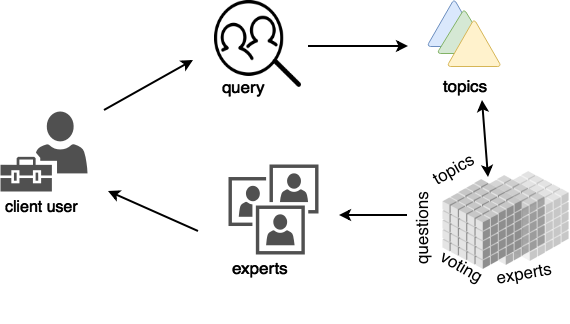}
  \caption{Work-flow of our proposed methodology: For a given input query, experts are output based on the detected topic of the query combined with our 4th order tensor, which contains latent information like topics, questions, voting, and experts.}
\label{fig:scr}
\end{figure}

We aim to recommend experts in multiple areas simultaneously.
In particular, we use the Stack Exchange networks dump, which contains various areas, to build up a multi-domain dataset. We propose group lasso~\cite{Kim:2010:TGL:3104322.3104392} that works on a relationship tree formed upon the natural structure of the SE network. The tree is used to guide the decomposition of 4th rank tensor data consisting of questions, topics, voting and expertise information. We additionally factorize selected matrices to provide additional latent information.

Our contributions in this work are as follows:
\begin{enumerate}
  \item We take the hierarchical relationship between participants and topics into account and build a model that combines tree-guided tensor decomposition and matrix factorization;
  \item We introduce the relationship tree group lasso to alleviate the data sparsity problem;
  \item We conduct experiments on real-world data and evaluate the proposed approach against state-of-the-art baselines.
\end{enumerate}

\section{Related Works}
Expert recommendation has been studied extensively in the past decade. Generally, skillfulness and resourcefulness of experts can assist users in making decisions more professionally and solving problems more effectively and efficiently. That is, making appropriate recommendations to users with the different requirement can be important.

The expert recommendation techniques apply to many areas, and different fields may require differently in methodologies to handle different situations. Baloga et al.~\cite{balog2009language} introduce a generative probabilistic framework for find experts in various enterprise data sources. Daud et al.~\cite{daud2010temporal} devise a Temporal-Expert-Topic model to capture both the semantic and dynamic expert information and to identify experts for different time periods. Fazelzarandi et al.~\cite{fazel2011expert} develop an expert recommendation system with utilizing the social networks analysis and multiple data source integration techniques. 
Wang et al.~\cite{wang2013expertrank} propose a model ExpertRank which take both document profile and authority of experts into consideration to perform better.  Huang et al.~\cite{8029777} take advantage of word embedding technology to rank experts both semantically and numerically. More relate works can be found in a survey by Wang et al.~\cite{wang2018survey}.

The works mentioned above mostly focus on recommend experts for organizations, enterprises or institutes. There is also some literature on recommending experts in Q\&A System, which is more related to our work. Kao et al.~\cite{kao2010expert} propose to incorporate user subject relevance, user reputation and authority of categories into expert finding system in Q\&A websites.
Riahi et al.~\cite{riahi2012finding} investigate two topic model namely Segmented Topic Model and Latent Dirichlet Allocation model to direct new questions in Stack-overflow to related experts. Ge et al.~\cite{ge2016taper} propose a personalized tensor-based method for expert recommendation by considering factors like geospatial, topical and preferences. Liu et al. in \cite{liu2015zhihurank} propose a method to rank user authority by exploiting interactions between users, which is aimed to avoid potential impacts of users with considerable social influences. They introduced topical similarities into link analysis to rank user authorities for each question. Latent Dirichlet allocation is applied to extract topics from both the questions and answers of users so that topical similarities between questions and answers can be measured, and then related users can be ranked by links.
Huna et al. found Q\&A communities often evaluate user reputation limited to the number of user activities\cite{huna2016exploiting}, regardless of efforts on creating high-quality contents. This causes inaccurate measurements in user expertise and their value. Inspired by former works, they calculate user reputations for asking and answering questions. The reputation results from the combination of the difficulty score of a question and the utility score for the question or answer. A utility score measures the distance between a score and the maximum score of the post, and the difficulty measures the times that a user spends on the question. The time spent on questions is normalized on each topic. Fang et al.~\cite{fang2016community} are well aware of the quantity of social information Q\&A website can provide, along with the importance of user-generated textual contents. Their idea to simultaneously model both social links and textual contents leads to the proposed framework named ``HSNL''(CQA via \textbf{H}eterogeneous \textbf{S}ocial \textbf{N}etwork \textbf{L}earning). The framework adopts random walk to exploit social information and build the heterogeneous social network, and a deep recurrent neural network was trained to give a text-based matching score for questions and answers.

Our proposed model builds on tensor decomposition, which has been applied to various fields such as neuroscience,  computer vision, and data mining~\cite{kolda2009tensor}. CANDECOMP/PARAFAC (CP) and Tucker decomposition are two effective ways to solve tensor decomposition problems. We adopt the former in this work. Tensor decomposition based recommender systems can also be found widespread in recent studies. Rendle et al.~\cite{rendle2009learning} introduce a tensor factorization based ranking approach for tag recommendation. They further improve the model by introducing pairwise interaction and significantly improve the optimization efficiency. Xiong et al.~\cite{xiong2010temporal} propose a probabilistic tensor decomposition model and regard the temporal dynamics as the third-dimension of the tensor.  Karatzoglou et al.~\cite{karatzoglou2010multiverse} offer a context-aware tensor decomposition model to integrate context information with collaborative filtering tightly. Hidas et al.~\cite{hidasi2012fast} investigate approach which combines implicit feedback with context-aware decomposition.  Bhargava et al.~\cite{bhargava2015and} present a tensor decomposition-based approach to model the influence of multi-dimensional data sources. Yao et al.~\cite{Yao:2015:CPR:2766462.2767794} decompose tensor with contextual regularization to recommend location points of interest.

\section{Methodology}
\label{sec:metho}

 CANDECOMP/PARAFAC Tensor Decomposition, or CP Decomposition, is discovered by Kiers and M\"ocks independently\cite{kolda2009tensor}. For a Rank-$R$ size-$N$ tensor $\mathcal{X}$ ($R\in\mathbb{N}$), let $U_1\in\mathbb{R}^{I_1\times{R}}, U_2\in\mathbb{R}^{I_2\times{R}}, ..., U_R\in\mathbb{R}^{I_N\times{R}}$, we have the decomposition:

\begin{equation}
  \mathcal{X} \approx \sum_{r=1}^R {U_1}_{i_1r}{U_2}_{i_2r}\cdots{U_R}_{i_Nr}
\end{equation}

While multiple methods can do tensor decomposition, the most common and effective one shall be the alternating least squares(ALS)\cite{kolda2009tensor}.

\subsection{Relationship Tree Modelling}

Our data is naturally divided into subsites, topics, and posts, as shown in Figure~\ref{fig:ht}. This decomposition forms a tree, with subsites on top, and posts as leaves. As our tensor models the expertise information based on user activities, this tree reserves the relationships of entities. We illustarte the contruction of the tree as follows.

\begin{figure}[ht!]
\centering
\includegraphics[width=0.6\linewidth]{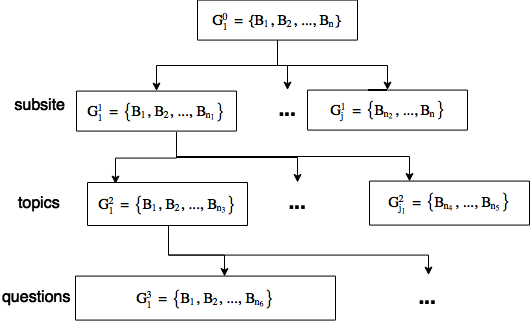}
\label{fig:tree_model}
\caption{An example of modeled tree representation of hierarchical relationship}
\end{figure}

Given the tree $\mathcal{T}$, we assume that the $i$-th level of $\mathcal{T}$ has $n_i$ nodes and organized as $\mathcal{T}_i = \{ G_1^i, G_2^i,..., G_{n_i}^i\}$. And so, a group $G_v$ where node $v \in V$ is in the tree, and all all leaves under $v$ are in $G_v$. Now we can define a tree-structured regulation as

\begin{equation}
Weight(\mathbf{U}_1)=\frac{\lambda_W}{2}\sum_{k=1}^{J}\omega_j^i\|\mathbf{U}_{1_k}\|_2^2 \quad \text{(} \mathbf{U}_{1_k} \in G_j^i \text{)}
\end{equation}

This inspired from Moreau-Yosida regularization, and here $\lambda_W$ is the Moreau-Yoshida regulation parameter for tree $\mathcal{T}$, $\|\cdot\|$ denotes Euclide  an norm, $\mathbf{U}_{1_k}$ is a vector of $\mathbf{U}_1$, where $\mathbf{U}_1$ is the first factor matrix of the tensor $\mathcal{X}$, which corresponding to a question post and detailed explaination can be found in the following subsection. Additionally, $\omega_j^i$ is set by following Kim's approach\cite{kim2010tree} and it means a pre-set weight for $j$-th node at level $i$. $\omega_j^i$ can be obtained by setting two variables summed up to 1, i.e. $s_j^i$ for the weight of independent relevant covariates selecting and $g_j^i$ for group relevant covariates selecting. We have:

\begin{equation}
\sum_{i}^{d}\sum_{j}^{n}\omega_i^j\|\mathbf{U}_{1_{G_j^i}}\|_2 = \lambda\omega_0^j
\end{equation}

where
\begin{equation}
  \omega_i^j = 
\begin{cases}
 s_j^i \cdot \sum_{c_p^q \in \text{Child}(v_j^i)}|\omega_p^q| + g_j^i \cdot \|\mathbf{U}_{1_{G_j^i}}\|_2 &v_j^i \text{is a internal node},\cr
|\mathbf{U}_{1_{G_j^i}}| & v_j^i \text{is a leaf node}.
\end{cases}
\end{equation}

\subsection{Proposed Model}

\begin{figure}
\centering
\begin{minipage}{0.45\columnwidth}
\includegraphics[width=\linewidth]{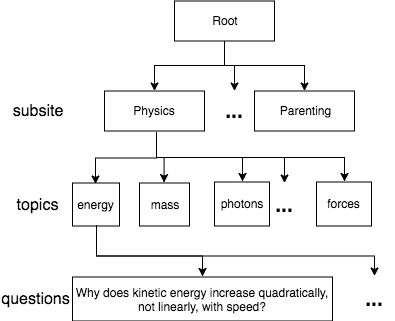}
\caption{Tree representation of hierarchical entity relationship}
\label{fig:ht}
\end{minipage}
\begin{minipage}{0.45\columnwidth}
\includegraphics[width=\linewidth]{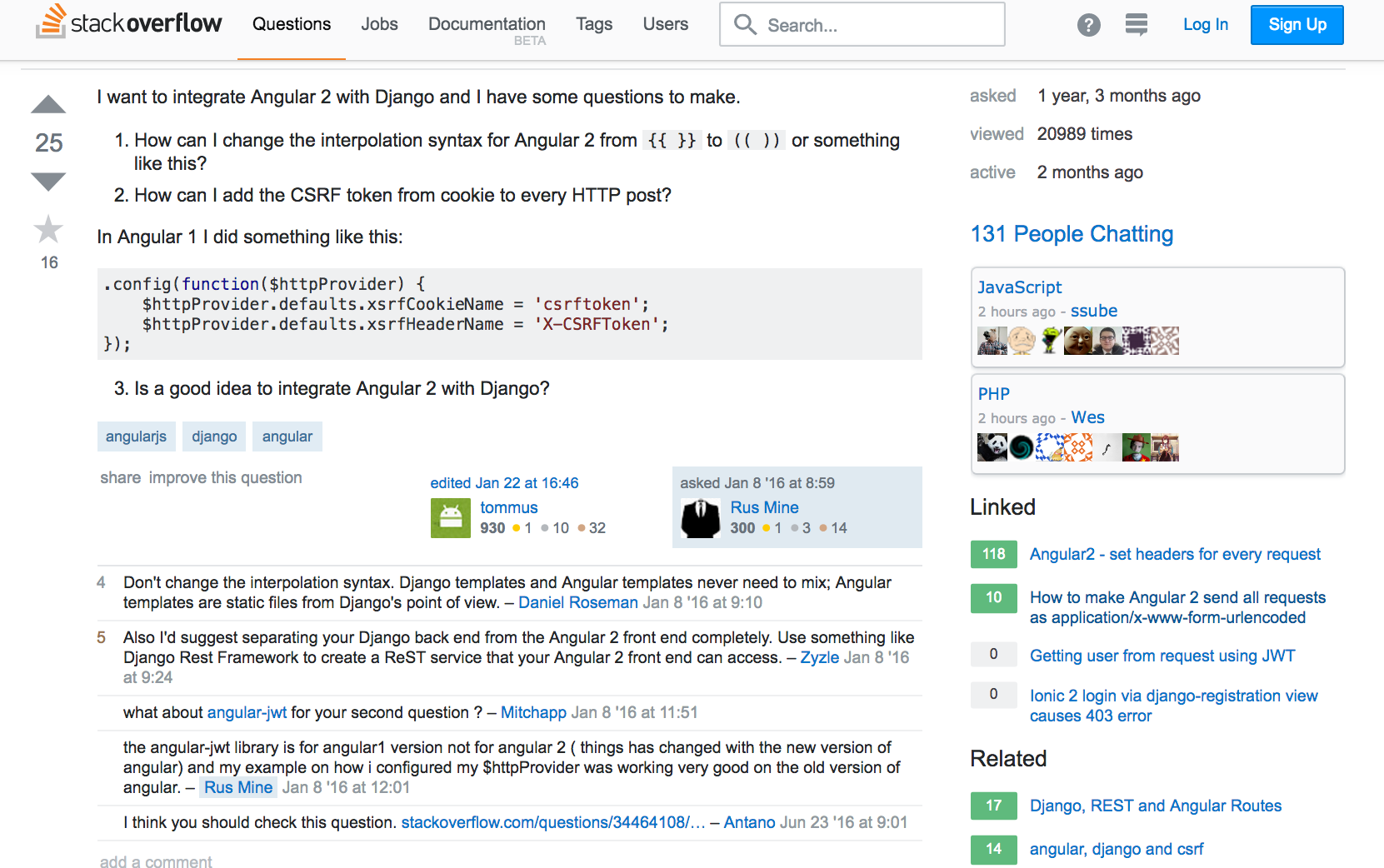}
\caption{An example of Stack Overflow post( postId:34672987), here demonstrates a question with its description and comments, along with score of the question.}
\label{fig:seeg}
\end{minipage}
\end{figure}

Our dataset is obtained naturally categorized by their subdomains, which we call it ``subsites'' here. Additionally, in each subsite, we can find tags in every post, and such information is often an indicator of the post's topics.
Accordingly, after gathering those data, we can build a tree to represent such hierarchical information( shown in Figure \ref{fig:ht}).

All Stack Exchange subsites share the same structure. That means, in all this subsites, answerers may propose multiple answers and questioners can adopt only one answer for each question. Also, both question and answers can be commented and voted, and the difference between vote-ups or vote-downs on each question is calculated into a score. Figure \ref{fig:seeg} show an example.

\begin{figure}[ht]
\centering
\includegraphics[width=0.8\linewidth]{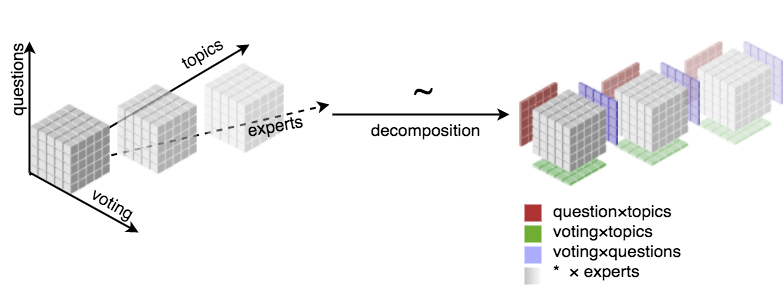}
\hspace{2em}
\caption{Proposed decomposition}
\label{fig:tensor}
\end{figure}

\begin{table}[ht]
\centering
\caption{Symbol table}
\label{tab:symbol}
\setlength\tabcolsep{5pt}
\begin{tabular}{ll}
\toprule
  Symbol & Description\\\midrule
 $\mathcal{X} \in \mathbb{R}^{I\times{J}\times{K}\times{L}}$
    &\makecell[l]{
      a 4th-order-tensor,\\
      $I, J, J, L$ accordingly is the number of \\
      \textbf{Question}, \textbf{Topic}, \textbf{Voting}\\
      and \textbf{Expert}
    }\\
&\\
 \makecell[l]{
   $\mathbf{U_1} \in \mathbb{R}^{I\times{R}}$\\
   $\mathbf{U_2} \in \mathbb{R}^{J\times{R}}$\\
   $\mathbf{U_3} \in \mathbb{R}^{X\times{R}}$\\
   $\mathbf{U_4} \in \mathbb{R}^{L\times{R}}$
 }
    &factor matrices of tensor $\mathcal{X}$ \\
&\\
 $M \in \mathbb{R}^{X\times{Z}}$
    &\makecell[l]{
      $\textbf{subsite}\times\textbf{answerer}$ matrix\\
      where $X,Z$ are the number of \textbf{subsite} \\
      and \textbf{answerer}
    }\\
&\\
$N \in \mathbb{R}^{Y\times{Z}}$
    &\makecell[l]{
      $\textbf{topic}\times\textbf{answerer}$ matrix\\
      where $Y$ are the number of \textbf{topic}
    }\\
&\\
 $\mathcal{T}_i = \{ G_1^i, G_2^i,..., G_{n_i}^i\}$
    & \makecell[l]{
      set of node in the $i$-th level of tree $\mathcal{T}$\\
      $G_{n_i}$ is the $n_i$-th node in the level
    }\\
\bottomrule
\end{tabular}
\end{table}

Instead of the simple score-user matrix based recommendation, we propose a tensor-decomposition based tree-guided method, based on the basic idea of Tree-Guided Sparse Learning\cite{jenatton2010proximal}.
\begin{enumerate}
  \item A 4th-order-tensor, $\textbf{Question}\times\textbf{Topic}\times\textbf{Voting}\times\textbf{Expert}$. Shown in Figure \ref{fig:tensor}, we denoted it as $\mathcal{X} \in \mathbb{R}^{I\times{J}\times{K}\times{L}}$, where $I$ is the number of questions, $J$ is the the number of Topics, $K$ is the number of voting of question towards questioners, $L$ is the expert users and the value of the tensor is the number of expertise evaluation criterion. With limited users participated in certain domains, it is believed that the tensor is very sparse. Additionally we denote $\mathbf{U_1} \in \mathbb{R}^{I\times{R}}, \mathbf{U_2} \in \mathbb{R}^{J\times{R}}, \mathbf{U_3} \in \mathbb{R}^{X\times{R}}, \mathbf{U_4} \in \mathbb{R}^{L\times{R}}$ as factor matrices of tensor $\mathcal{X}$.

  \item A $\textbf{subsite}\times\textbf{answerer}$ matrix. We denoted this as $M \in \mathbb{R}^{X\times{Z}}$, where if answerer $z$ appears in subsite $x$, $M_{x,z} = 1$ else $M_{x,z} = 0$.

  \item A $\textbf{topics}\times\textbf{answerer}$ matrix. We denoted this as $N \in \mathbb{R}^{Y\times{Z}}$, similarly here, when answerer $z$ appears in topic $y$, $M_{y,z} = 1$ else $M_{y,z} = 0$.

  \item Hierarchical relationship tree $\mathcal{T}$ of depth $d$. Due to the isolation of subsites and their topics, our data show clearly a  structured sparsity. Thus, we can utilize tree-guided group lasso in our model. That is, besides above two supplement matrices, we also use the tree shown in Figure\ref{fig:ht} to guide the learning.
\end{enumerate}

After modeling the data, we  apply CANDECOMP/PARAFAC (CP) tensor decomposition to factorize the tensor and solve the tree-structured regression with group lasso.

\begin{algorithm}[ht]
  \begin{algorithmic}[1]
  \renewcommand{\algorithmicrequire}{\textbf{Input:}}
  \renewcommand{\algorithmicensure}{\textbf{Output:}}
  \REQUIRE $\mathcal{X}, \mathit{R}$
  \ENSURE $\lambda, \mathbf{A}^{(1)}, \mathbf{A}^{(2)}, ..., \mathbf{A}^{(n)}$
  \\\hspace{1em}
  \\\textit{\bf Algorithm} $CP-ALS(\mathcal{X}, \mathit{R})$
  \\\text{initialize $\mathbf{A}^{(n)} \in \mathbb{R}^{\mathit{I_n\times{R}}}$ for $n = 1, 2, ..., N$}
  \FOR {$n = 1, 2, ..., N$}
    \STATE $\mathbf{V} \leftarrow \mathbf{A}^{(1)\top}\mathbf{A}^{(1)} \ast ... \ast \mathbf{A}^{(n-1)\top}\mathbf{A}^{(n-1)} \ast \mathbf{A}^{(n+1)\top}\mathbf{A}^{(n+1)} \ast$\\$ ... \ast \mathbf{A}^{(N)\top}\mathbf{A}^{(N)}$
    \STATE $\mathbf{A}^{(n)} \leftarrow \mathbf{X}^{n}(\mathbf{A}^{(n)}\odot ... \odot \mathbf{A}^{(n+1)} \odot \mathbf{A}^{(n-1)} \odot ... \odot \mathbf{A}^{(1)})\mathbf{V}^{\ast}$
    \STATE normalize columns of $\mathbf{A^{(n)}}$ and store norms as $\lambda$
    \IF {fit stops improve \OR iteration reach threshold}
      \STATE \text{\bf break}
    \ENDIF
  \ENDFOR
  \RETURN $\lambda, \mathbf{A}^{(1)}, \mathbf{A}^{(2)}, ..., \mathbf{A}^{(n)}$
  \end{algorithmic}
  \caption{CP Decomposition via Alternating Least Squares, where $N$-th order tensor $\mathcal{X}$ of size $\mathit{I}_1\times\mathit{I}_2\times ... \times\mathit{I}_N$ is decomposite into $R$ components}
  \label{alg:cp_als}
  \end{algorithm}

First, we decompose the 4th-order tensor with regulation by Alternating Least Square (ALS) as follows:
\begin{equation}
\label{eq:tensor_dec}
\begin{aligned}
Tensor(\mathbf{U_1}, \mathbf{U_2}, \mathbf{U_3}, \mathbf{U_4})&=\frac{1}{2}
\|\mathcal{X}-{\llbracket \mathbf{U_1}, \mathbf{U_2}, \mathbf{U_3}, \mathbf{U_4} \rrbracket}\|_F^2 \\
&+ \frac{\lambda_{\mathcal{X}}}{2}
(\|\mathbf{U_1}\|_F^2 + \|\mathbf{U_2}\|_F^2 + \|\mathbf{U_3}\|_F^2 + \|\mathbf{U_4}\|_F^2)
\end{aligned}
\end{equation}

Then, we can have the aforementioned 2 matrices decompose as :
\begin{equation}
Networks(\mathbf{S}, \mathbf{A})=\frac{1}{2}\|\mathbf{M}_{site} - \mathbf{SA}^T\|_F^2 +
 \frac{\lambda_S}{2}(\|\mathbf{S}\|_F^2 + \|\mathbf{A}\|_F^2)
\end{equation}

\begin{equation}
Topic(\mathbf{T}, \mathbf{A})=\frac{1}{2}\|\mathbf{M}_{topic} - \mathbf{TA}^T\|_F^2 +
 \frac{\lambda_T}{2}(\|\mathbf{T}\|_F^2 + \|\mathbf{A}\|_F^2)
\end{equation}

Since each subsite $S_j$ contains a group of questions $\mathbf{U}_{1_{j}}$, we expect $S_j$ to be similar to the average $\mathbf{U}_{1_{j}}$, which can be solved as a regulation:
\begin{equation}
Site(\mathbf{S}, \mathbf{U_1})=\frac{\lambda_S}{2}\sum_{j=1}^{U}\|\mathbf{S}_j-\frac{1}{G_j^1}\sum_{\mathbf{U}_{1_k} \in G_j^1}\mathbf{U}_{1_k}\|_2^2
\end{equation}

By combining those objectives and regulations, we have the following objective function:

\begin{equation}
\begin{aligned}
f(\mathbf{U_1}, \mathbf{U_2}, \mathbf{U_3}, \mathbf{U_4}, \mathbf{S}, \mathbf{A}, \mathbf{T})&=
Tensor(\mathbf{U_1}, \mathbf{U_2}, \mathbf{U_3}, \mathbf{U_4})\\
&+Weight(\mathbf{U}_1) + Networks(\mathbf{S}, \mathbf{A})\\
&+ Topic(\mathbf{T}, \mathbf{A}) + Site(\mathbf{S}, \mathbf{U_2})
\end{aligned}
\end{equation}

Equation~\ref{eq:tensor_dec} follows the CANDECOMP/PARAFAC Decomposition, accomplished by the ALS algorithm (see Algorithm~\ref{alg:cp_als}), which is a popular way to decompose a tensor.

{\bf Computational Complexity Analysis.} The time complexity of the above decomposition includes two parts.
The first concerns initializing the set of $\mathbf{A}^{(n)}$s. We note the average of the dimension of our tensor as $D$, which we use to represent the size of the tensor as $\mathbf{D}^{N}$. The initialization is a traverse of $\mathbf{A}^{(n)}$s and has a time complexity of $\mathcal{O}(NDR)$.
Assuming that we use index flip to implement the matrix transpose, its time complexity is $\mathcal{O}(1)$. Thus, the total time complexity on $N$ loops is $\mathcal{O}((NDR)^2 + N^2DR)$ time.
Combining the two steps, we now have the time complexity of the algorithm as $\mathcal{O}((NDR)^2)$.

\section{Experiments and Evaluation}
\label{sec:expr}
In this section, we report our experiments to evaluate our proposed approach. We first briefly introduce our dataset and the evaluation metrics, and then present the results analysis and evaluation. 

Until now, there is no ``gold standard'' to evaluate our approach regarding expert recommendation, to the best of our knowledge. Also, it is difficult to judgment user's expertise manually due to the large-scale data (e.g., our test data contains more than 2 million users and nearly 20 million voting activities on 5 million posts) and the lack of ranking information in the dataset---the reputation scores of users in Stack Exchange systems are computed globally, which cannot be utilized to evaluate individual's ability in specific domains or topics.

Similar to Huna et al. \cite{huna2016exploiting}, we calculate the reputation score of each user by topics, according to the rules adopted by Stack Exchange\footnote{\url{https://stackoverflow.com/help/whats-reputation}}. We simplify the rule by removing bounty-related and edition-related reputation differences. Table~\ref{tab:reputuation} summarizes the simplification results. A rank can be established based on the built-in reputation scores of users, following the approach proposed by Huna et al.\cite{huna2016exploiting}. The rank serves as a baseline for comparative performance evaluation. Given the lack of a standard to measure verifiable expertise of users, we adopt this idea and conduct comparison experiments.

\begin{table}[ht]
  \centering
  \caption{Adopetd reputation rules}
  \label{tab:reputuation}
  \begin{tabular}{rp{3cm}<{\centering}} 
    \toprule
  activity                       & reputation gaines \\ \midrule
  Answer is upvoted              & +10               \\
  Question is upvoted            & +5                \\
  Answer/question is downvoted   & -2                \\
  Downvote an answer             & -1                \\
  Answer is acceped              & +15               \\ \bottomrule
  \end{tabular}
  \end{table}

\subsection{Dataset and Experiment Settings}

\begin{table}[ht!]
  \centering
  \caption{Selected statistics profiles of experiment dataset}
  \label{tab:data_profile}
  \begin{tabular}{>{\raggedleft\arraybackslash}p{2cm}p{1.5cm}<{\centering}p{1.5cm}<{\centering}p{1.5cm}<{\centering}p{1.5cm}<{\centering}}
  \toprule
           & \textbf{\# of Users} & \textbf{\# of Posts} & \textbf{\# of Tags} & \textbf{\# of Votes}\\ \midrule
  \textbf{apple}       & 153360 & 202239  & 1048 & 720540  \\ 
  \textbf{askUbuntu}   & 420227 & 598530  & 3022 & 2543467 \\ 
  \textbf{gis}         & 63977  & 179507  & 2221 & 573263  \\ 
  \textbf{math}        & 315792 & 1807772 & 1518 & 6046107 \\ 
  \textbf{physics}     & 95485  & 234583  & 876  & 1055850 \\ 
  \textbf{serverFault} & 302850 & 645711  & 3514 & 2048746 \\ 
  \textbf{stat}        & 111974 & 195038  & 1331 & 782689  \\ 
  \textbf{superuser}   & 500264 & 859690  & 5190 & 3281616 \\ 
  \textbf{unix}        & 188934 & 284114  & 2438 & 1276409 \\ \bottomrule
  \end{tabular}
  \end{table}

\subsubsection{Dataset} As mentioned above, the Stack Exchange Networks includes 98 subsites and massive data. We identified 14,220,976 users, 46,575,393 posts, 178,575 tags, and 178,184,014 votes. Computing at such a scale can be challenging to any existing systems. Thus, in this work, we conducted experiments on several reasonably selected subsets, which contains a feasible yet still decent volume of data.

Note that, our method is a tree-guided tensor decomposition approach, where the tree models the hierarchical entity relationships including topics information. To keep the variance of the topics, we generate our testing subsets from sereval independent subsites. 
These subsites are named as 
\emph{  ``apple'', ``math'', ``stats'', ``askubuntu'', ``physics'', ``superuser'', ``gis'',``serverfault'', and ``unix''.}
Some selected statistics profiles can be found at Table~\ref{tab:data_profile}.

Due to the massive scale of our data source and its high degree of sparseness, a random sampling could end up output posts with an enormous number of unrelated users and topics. Hence, we first sample randomly to select a subset of users and then enumerations on posts tags and voting are performed. This ensures the selected posts and votes are all related to the sampled users.

\subsection{Results Analysis and Evaluation}

\subsubsection{Evaluation Metrics}
\begin{itemize}
  \item {\bf Precision@$k$} Precision@$k$ is one of standard evaluation metrics in information retrieval tasks and recommender systems. It is defined to calculate the proportion of retrieved items in the top-$k$ set that are relevant. Here our frameworks return a list of users so that the Precision@$k$ can be calculated as follows:
  \begin{equation*}
    P@k=\frac{|\{relevant\_top-k\_users\}\cap\{retrived\_top-k\_users\}|}{|\{retrived\_top-k\_users\}|}
  \end{equation*}   
  \item {\bf MRR} The Mean Reciprocal Rank is a statistic measure for evaluating response orderly to a list, which here is average of reciprocal ranks for all tested questions:
  \begin{equation*}
    MRR=\frac{1}{|Q|}\sum{i=1}{|Q|}\frac{1}{Rank_i}
  \end{equation*}
\end{itemize}

\subsubsection{Compared methods}
\begin{itemize}
  \item {\bf Baselines} Apart from the reputation value calculated by Stack Exchange rules mentioned earlier in Table~\ref{tab:reputuation}, it also can be found that some baselines are also often used apart from reputation value. Namely, lists generated by rank by "Best Answer Ratio" of users and rank by "Number of Answers" produced by users.  
  \item {\bf MF-BPR\cite{rendle2009bpr}} Rendel et al. introduce pairwise BPR ranking loss into standard Matrix Factorization models. It is specifically designed to optimize ranking problems.
  \item {\bf Zhang et al.\cite{zhang2007expertise}}, Z-Score by Zhang et al., is a well-known reputation measure, despite their original work is a PageRank based system and is not aimed at measurements. This feature-based score can be resolved by $q$ the number of questions a user asked and $ a$, the number of answers the user posted. That is,
  \begin{equation*}
    Z-Score=\frac{a-q}{\sqrt{a+q}}  
  \end{equation*}
  \item {\bf ConvNCF\cite{heouter}} Outer Product-based Neural Collaborative Filtering, a multi-layer neural network architecture based collaborative filtering method. it use an outer product to find out the pairwise correlations between the dimensions of the embedding space.
\end{itemize}

\subsubsection{Results Analysis}
\begin{figure}[ht!]
  \centering    
  \includegraphics[width=0.4\linewidth]{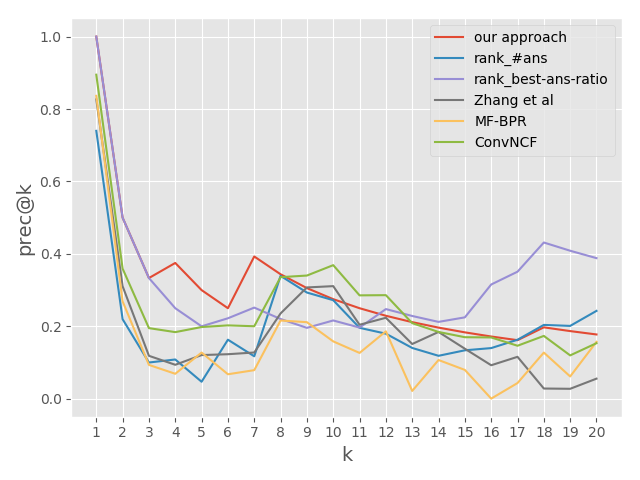}
  \includegraphics[width=0.4\linewidth]{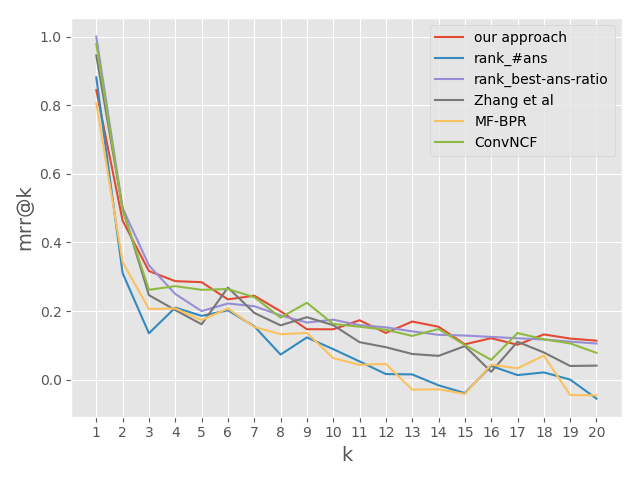}
\caption{Preformance comparison of our approach to others, tested with 250 users and their historical data}
\label{fig:perf}
\end{figure}

\begin{figure}[ht!]
  \centering
  \includegraphics[width=0.4\linewidth]{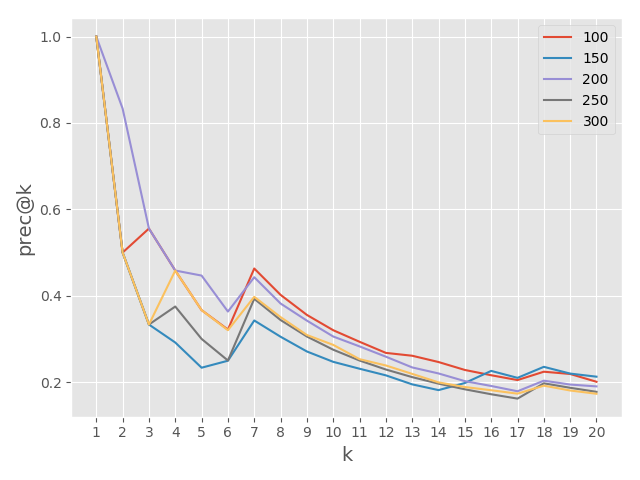}
  \includegraphics[width=0.4\linewidth]{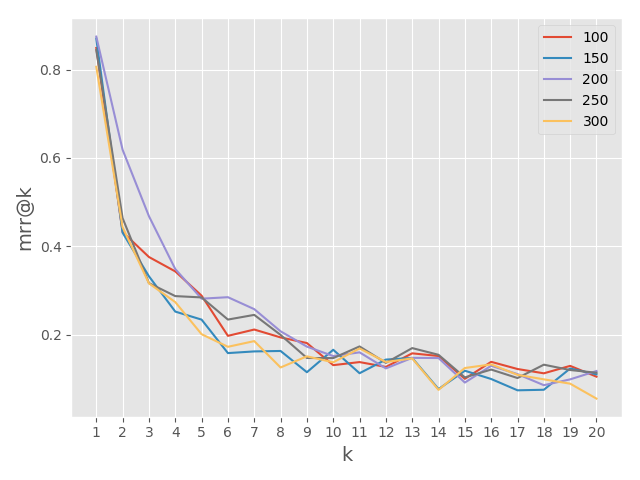}
\caption{Precision and MRR of tests at various number of users}
\label{fig:stability}
\end{figure}

\begin{figure}[ht!]
  \centering
  \includegraphics[width=0.8\linewidth]{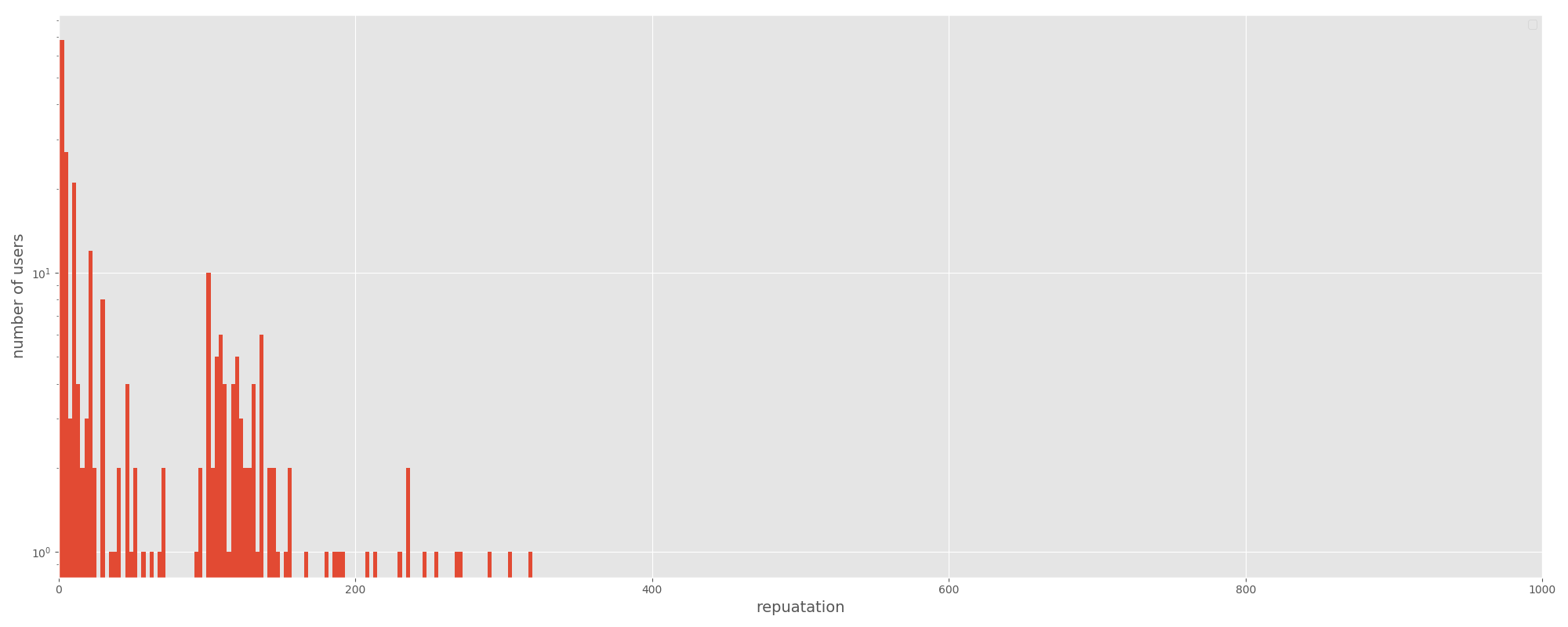}
\caption{Distribution of reputation of users in our dataset}
\label{fig:dist}
\end{figure}

Figure~\ref{fig:perf} shows the evaluation results with respect to the Precision and MRR of different methods, where precision measures the ability to find experts and MRR the performance of outputting list of experts in correct order. We observed that our approach generally outperformed other tested approaches, although some other approaches produces more accurate list when the length of the requested list is no more than 3, and this can be claimed less likely to be practical. Our approach yielded better ranks in most cases except some case where very short lists were requested. Yet, It can be argued, in real life applications, a the list of approximately 10 or more experts is largely sensible and our approach will have substantial better performance. Also interestingly, here we can see both precision and MRR decreases by the increase of $K$, which differs from our experience of previous work. And a further look at the distribution of reputation in our tested data reveals it actually sensible, as we can see in Figure~\ref{fig:dist}, the distribution of users' reputation is considerably uneven, given very few people high have reputation, which are our goal of output, and most people in the dataset are reputed at value 1. Additionally, to assess the stability of our approach, we conducted tests with various size of input data, ranging from 100 users to 300 users. Besides acceptable fluctuations, the results demonstrate our approach performs relatively stable, both in accuracy and quality. 

\section{Conclusion}
\label{sec:concl}
In this paper, we have proposed a framework to identify experts across different collaborative networks.
The framework use tree-guided tensor decomposition to exploit insights from Q\&A networks. In particular, we decomposite a 4th rank tensor with tree-guided lasso and matrix factorization to exploit the topic information from a collection of Q\&A websites in Stack Exchange Networks to alleviate the data sparsity issue. The 4th rank tensor model of the data ensures to keep as much as information as needed, which confirmed by experiments and evaluation. Due to the lack of ``Gold Standard'', we compared our approach with baselines accordingly to the rank by the reputation score calculated by Stack Exchange built-in approaches on each topic.
The comparison results demonstrate the feasibility of our approach. The proposed approach can be applied to broader scenarios such as finding the most appropriate person to consult on some specific problems for individuals, or identifying the desired employees for enterprises. 

\section*{Acknowledgment}
 This research was undertaken with the assistance of resources and services from the National Computational Infrastructure (NCI), which is supported by the Australian Government.

\bibliographystyle{splncs04}

\begin{thebibliography}{10}
\providecommand{\url}[1]{\texttt{#1}}
\providecommand{\urlprefix}{URL }
\providecommand{\doi}[1]{https://doi.org/#1}

\bibitem{amatriain2009wisdom}
Amatriain, X., Lathia, N., Pujol, J.M., Kwak, H., Oliver, N.: The wisdom of the
  few: a collaborative filtering approach based on expert opinions from the
  web. In: Proceedings of the 32nd international ACM SIGIR conference on
  Research and development in information retrieval. pp. 532--539. ACM (2009)

\bibitem{balog2009language}
Balog, K., Azzopardi, L., de~Rijke, M.: A language modeling framework for
  expert finding. Information Processing \& Management  \textbf{45}(1),  1--19
  (2009)

\bibitem{bhargava2015and}
Bhargava, P., Phan, T., Zhou, J., Lee, J.: Who, what, when, and where:
  Multi-dimensional collaborative recommendations using tensor factorization on
  sparse user-generated data. In: Proceedings of the 24th International
  Conference on World Wide Web. pp. 130--140. ACM (2015)

\bibitem{daud2010temporal}
Daud, A., Li, J., Zhou, L., Muhammad, F.: Temporal expert finding through
  generalized time topic modeling. Knowledge-Based Systems  \textbf{23}(6),
  615--625 (2010)

\bibitem{fang2016community}
Fang, H., Wu, F., Zhao, Z., Duan, X., Zhuang, Y., Ester, M.: Community-based
  question answering via heterogeneous social network learning. In: Thirtieth
  AAAI Conference on Artificial Intelligence (2016)

\bibitem{fazel2011expert}
Fazel-Zarandi, M., Devlin, H.J., Huang, Y., Contractor, N.: Expert
  recommendation based on social drivers, social network analysis, and semantic
  data representation. In: Proceedings of the 2nd international workshop on
  information heterogeneity and fusion in recommender systems. pp. 41--48. ACM
  (2011)

\bibitem{ge2016taper}
Ge, H., Caverlee, J., Lu, H.: Taper: A contextual tensor-based approach for
  personalized expert recommendation. Proc. of RecSys  (2016)

\bibitem{heouter}
He, X., Du, X., Wang, X., Tian, F., Tang, J., Chua, T.S.: Outer product-based
  neural collaborative filtering

\bibitem{hidasi2012fast}
Hidasi, B., Tikk, D.: Fast als-based tensor factorization for context-aware
  recommendation from implicit feedback. Machine Learning and Knowledge
  Discovery in Databases pp. 67--82 (2012)

\bibitem{8029777}
Huang, C., Yao, L., Wang, X., Benatallah, B., Sheng, Q.Z.: Expert as a service:
  Software expert recommendation via knowledge domain embeddings in stack
  overflow. In: 2017 IEEE International Conference on Web Services (ICWS). pp.
  317--324 (June 2017). \doi{10.1109/ICWS.2017.122}

\bibitem{huna2016exploiting}
Huna, A., Srba, I., Bielikova, M.: Exploiting content quality and question
  difficulty in cqa reputation systems. In: International Conference and School
  on Network Science. pp. 68--81. Springer (2016)

\bibitem{jenatton2010proximal}
Jenatton, R., Mairal, J., Bach, F.R., Obozinski, G.R.: Proximal methods for
  sparse hierarchical dictionary learning. In: Proceedings of the 27th
  international conference on machine learning (ICML-10). pp. 487--494 (2010)

\bibitem{kao2010expert}
Kao, W.C., Liu, D.R., Wang, S.W.: Expert finding in question-answering
  websites: a novel hybrid approach. In: Proceedings of the 2010 ACM symposium
  on applied computing. pp. 867--871. ACM (2010)

\bibitem{karatzoglou2010multiverse}
Karatzoglou, A., Amatriain, X., Baltrunas, L., Oliver, N.: Multiverse
  recommendation: n-dimensional tensor factorization for context-aware
  collaborative filtering. In: Proceedings of the fourth ACM conference on
  Recommender systems. pp. 79--86. ACM (2010)

\bibitem{Kim:2010:TGL:3104322.3104392}
Kim, S., Xing, E.P.: Tree-guided group lasso for multi-task regression with
  structured sparsity. In: Proceedings of the 27th International Conference on
  International Conference on Machine Learning. pp. 543--550. ICML'10,
  Omnipress, USA (2010),
  \url{http://dl.acm.org/citation.cfm?id=3104322.3104392}

\bibitem{kim2010tree}
Kim, S., Xing, E.P.: Tree-guided group lasso for multi-task regression with
  structured sparsity. In: Proceedings of the 27th International Conference on
  International Conference on Machine Learning. pp. 543--550. ICML'10,
  Omnipress, USA (2010),
  \url{http://dl.acm.org/citation.cfm?id=3104322.3104392}

\bibitem{kolda2009tensor}
Kolda, T.G., Bader, B.W.: Tensor decompositions and applications. SIAM review
  \textbf{51}(3),  455--500 (2009)

\bibitem{Liu2013IEP}
Liu, D.R., Chen, Y.H., Kao, W.C., Wang, H.W.: Integrating expert profile,
  reputation and link analysis for expert finding in question-answering
  websites. Inf. Process. Manage.  \textbf{49}(1),  312--329 (Jan 2013).
  \doi{10.1016/j.ipm.2012.07.002},
  \url{http://dx.doi.org/10.1016/j.ipm.2012.07.002}

\bibitem{liu2015zhihurank}
Liu, X., Ye, S., Li, X., Luo, Y., Rao, Y.: Zhihurank: A topic-sensitive expert
  finding algorithm in community question answering websites. In: International
  Conference on Web-Based Learning. pp. 165--173. Springer (2015)

\bibitem{rendle2009learning}
Rendle, S., Balby~Marinho, L., Nanopoulos, A., Schmidt-Thieme, L.: Learning
  optimal ranking with tensor factorization for tag recommendation. In:
  Proceedings of the 15th ACM SIGKDD international conference on Knowledge
  discovery and data mining. pp. 727--736. ACM (2009)

\bibitem{rendle2009bpr}
Rendle, S., Freudenthaler, C., Gantner, Z., Schmidt-Thieme, L.: Bpr: Bayesian
  personalized ranking from implicit feedback. In: Proceedings of the
  twenty-fifth conference on uncertainty in artificial intelligence. pp.
  452--461. AUAI Press (2009)

\bibitem{riahi2012finding}
Riahi, F., Zolaktaf, Z., Shafiei, M., Milios, E.: Finding expert users in
  community question answering. In: Proceedings of the 21st International
  Conference on World Wide Web. pp. 791--798. ACM (2012)

\bibitem{srba2016stack}
Srba, I., Bielikova, M.: Why is stack overflow failing? preserving
  sustainability in community question answering. IEEE Software
  \textbf{33}(4),  80--89 (2016)

\bibitem{wang2013expertrank}
Wang, G.A., Jiao, J., Abrahams, A.S., Fan, W., Zhang, Z.: Expertrank: A
  topic-aware expert finding algorithm for online knowledge communities.
  Decision Support Systems  \textbf{54}(3),  1442--1451 (2013)

\bibitem{wang2018survey}
Wang, X., Huang, C., Yao, L., Benatallah, B., Dong, M.: A survey on expert
  recommendation in community question answering. Journal of Computer Science
  and Technology  \textbf{33}(4),  625--653 (2018)

\bibitem{xiong2010temporal}
Xiong, L., Chen, X., Huang, T.K., Schneider, J., Carbonell, J.G.: Temporal
  collaborative filtering with bayesian probabilistic tensor factorization. In:
  Proceedings of the 2010 SIAM International Conference on Data Mining. pp.
  211--222. SIAM (2010)

\bibitem{Yao:2015:CPR:2766462.2767794}
Yao, L., Sheng, Q.Z., Qin, Y., Wang, X., Shemshadi, A., He, Q.: Context-aware
  point-of-interest recommendation using tensor factorization with social
  regularization. In: Proceedings of the 38th International ACM SIGIR
  Conference on Research and Development in Information Retrieval. pp.
  1007--1010. SIGIR '15, ACM, New York, NY, USA (2015).
  \doi{10.1145/2766462.2767794},
  \url{http://doi.acm.org/10.1145/2766462.2767794}

\bibitem{yao2018collaborative}
Yao, L., Sheng, Q.Z., Wang, X., Zhang, W.E., Qin, Y.: Collaborative location
  recommendation by integrating multi-dimensional contextual information. ACM
  Transactions on Internet Technology (TOIT)  \textbf{18}(3), ~32 (2018)

\bibitem{zhang2007expertise}
Zhang, J., Ackerman, M.S., Adamic, L.: Expertise networks in online
  communities: structure and algorithms. In: Proceedings of the 16th
  international conference on World Wide Web. pp. 221--230. ACM (2007)

\end{thebibliography}

\end{document}